\documentclass{JHEP3}
\usepackage{latexsym}
\def\AdSs5{$AdS_5$}
\def\AdS5s5{$AdS_5 \times S^5$}

\newcommand{\eg}{{\it e.g.~}}
\newcommand{\ie}{{\it i.e.~}}

\newcommand{\be}{\begin{equation}}
\newcommand{\ee}{\end{equation}}
\newcommand{\ba}{\begin{eqnarray}}
\newcommand{\ea}{\end{eqnarray}}

\newbox\SlashedBox
\def\fs#1{\setbox\SlashedBox=\hbox{#1}
\hbox to
0pt{\hbox to 1\wd\SlashedBox{\hfil/\hfil}\hss}{#1}}
\def\hboxtosizeof#1#2{\setbox\SlashedBox=\hbox{#1}
\hbox to
1\wd\SlashedBox{#2}}

\def\ms#1{\setbox\SlashedBox=\hbox{$#1$}
\hbox to 0pt{\hbox to
1\wd\SlashedBox{\hfil/\hfil}\hss}#1}

\def\t2{\tau_2}
\def\AdSS5{$AdS_5$}
\def\AdS5s5{$AdS_5
\times S^5$}

\def\IZ{\relax\ifmmode\mathchoice {\hbox{\cmss Z\kern-.4em
Z}}{\hbox{\cmss Z\kern-.4em Z}} {\lower.9pt\hbox{\cmsss Z\kern-.4em
Z}} {\lower1.2pt\hbox{\cmsss Z\kern-.4em Z}}\else{\cmss Z\kern-.4em
Z}\fi}

\def\g{\gamma}

\def\calT{{\cal T}}

\def\c1{{\chi^1}}

\def\N4{{\cal N}=4}
\def\half{{1\over 2}}
\def\nn{\nonumber}

\def\ie{{\it i.e.}}
\def\half{{{1 \over 2}}}

\def\tS{\tilde{S}}

\def\hm{{\widehat m}}

\def\hom{{\widehat\omega}}
\def\pmb#1{\setbox0=\hbox{#1}%
 \kern-.025em\copy0\kern-\wd0
 \kern.05em\copy0\kern-\wd0
 \kern-.025em\raise.0433em\box0 }
\font\cmss=cmss10
\font\cmsss=cmss10 at 7pt
\def\rlx{\relax\leavevmode}
\def\Zop{\rlx\leavevmode\ifmmode\mathchoice{\hbox{\cmss Z\kern-.4em Z}}
 {\hbox{\cmss Z\kern-.4em Z}}{\lower.9pt\hbox{\cmsss Z\kern-.36em Z}}
 {\lower1.2pt\hbox{\cmsss Z\kern-.36em Z}}\else{\cmss Z\kern-.4em
 Z}\fi}

\def\bbbone {{\mathchoice {\rm 1\mskip-4mu l} {\rm 1\mskip-4mu l}
{\rm 1\mskip-4.5mu l} {\rm 1\mskip-5mu l}}}
\def\pmb#1{\setbox0=\hbox{#1}%
 \kern-.025em\copy0\kern-\wd0
 \kern.05em\copy0\kern-\wd0
 \kern-.025em\raise.0433em\box0 }
\font\cmss=cmss10
\font\cmsss=cmss10 at 7pt
\def\rlx{\relax\leavevmode}
\def\Cop{\mathbb{C}}
\def\Rop{\relax{\rm I\kern-.18em R}}
\def\Nop{\relax{\rm I\kern-.18em N}}
\def\Pop{\relax{\rm I\kern-.18em P}}

\def\Zop{\rlx\leavevmode\ifmmode\mathchoice{\hbox{\cmss Z\kern-.4em Z}}
 {\hbox{\cmss Z\kern-.4em Z}}{\lower.9pt\hbox{\cmsss Z\kern-.36em Z}}
 {\lower1.2pt\hbox{\cmsss Z\kern-.36em Z}}\else{\cmss Z\kern-.4em
 Z}\fi}

\title{Oblique and curved D-branes in IIB plane-wave string theory}
\author{Matthias R. Gaberdiel\footnote{Address from 1 July 2003:
Institute for Theoretical Physics, ETH H\"onggerberg, CH-8093
Z\"urich, Switzerland.}
\\ {\it Department of Mathematics, \\
King's College London, Strand, London WC2R 2LS,
UK \\ \email{mrg@mth.kcl.ac.uk}\\ } }\author{Michael B. Green, Sakura
Sch\"afer-Nameki and Aninda Sinha\\ {\it Department of Applied
Mathematics and
Theoretical Physics,\\ Wilberforce Road, Cambridge CB3 0WA,
UK} \\ \email{M.B.Green, S.Schafer-Nameki, A.Sinha@damtp.cam.ac.uk}}

\abstract{Oblique $Dp$-branes in the maximally supersymmetric type IIB
plane-wave background are constructed in terms of boundary states, as
well as from the open string point of view. These $Dp$-branes, whose
existence was anticipated by Hikida and Yamaguchi from general
supersymmetry arguments, have an isometry that is a subgroup of the
diagonal $SO(4)$ symmetry of the background.  The oblique $D3$-brane
is found to preserve four dynamical and four kinematical
supersymmetries while the oblique $D5$-brane preserves one half of
both the dynamical and kinematical supersymmetries. We also discuss
the open-string boundary conditions for curved $D7$- and $D5$-branes,
and analyze their supersymmetry.}

\preprint{KCL-MTH-03-06 \\ DAMTP-2003-51}

\begin{document}

\section{Introduction}

The study of string theory in $pp$-wave backgrounds provides an
intriguing arena for extending the AdS/CFT correspondence to a
situation in which excited string states can explicitly be related to
composite Yang--Mills operators \cite{maldetal}.  The maximally
supersymmetric type IIB plane-wave background can be obtained from
$AdS_5 \times S^5$ by an infinite boost along a great circle of the
five-sphere -- the Penrose limit \cite{hulletal}.  Some supersymmetric
$D$-branes of the plane-wave string theory can be obtained by taking
the Penrose limits of BPS branes of the $AdS_5\times S^5$ theory.
However, others originate from non-supersymmetric branes of the
$AdS_5\times S^5$ theory and only become supersymmetric in the Penrose
limit.

The classification of $D$-branes in $pp$-wave backgrounds is
incomplete.  Two classes of $D$-branes have been explicitly
constructed in the maximally supersymmetric case in
\cite{bp,dp,st1,bgg,st2,gg,st3}, making use of a boundary state
closed-string formalism as well as by describing the open-string
states.  Such branes are characterised by a gluing matrix, $M$, that
describes how the left- and right-moving fermionic zero-modes of the
closed string are related at the boundary.  The orthogonal matrix $M$
is determined in terms of the geometry of the world-volume; if we
label by ${\cal N}$ an orthonormal basis for the world-volume
directions of the brane, then $M$ is given by
\be\label{Mdef}
M = \prod_{J\in{\cal N}} \gamma^J \,,
\ee
where $\gamma^J_{a\dot b}$ are $SO(8)$ gamma matrices, which couple
inequivalent spinors (with dotted and undotted indices). The two
classes of $D$-branes are:
\medskip

\noindent {\bf $\bullet$  Class I.} The branes of this class are
characterised by the condition
\be\label{pmpmI}
\left(\Pi M\Pi M\right)_{ab}= -\delta_{ab} \,,
\ee
where $\Pi = \gamma^1\gamma^2\gamma^3\gamma^4$.\footnote{We will
also define $\widehat \Pi = \gamma^5\gamma^6\gamma^7\gamma^8$. A
dotted spinor will be chosen to have positive $SO(8)$ chirality so
that $\left(\Pi \widehat \Pi\right)_{\dot a \dot b} =
+\delta_{\dot a \dot b}$ and
$\left(\Pi \widehat \Pi\right)_{a b} = - \delta_{ab}$.}
All of these branes preserve eight dynamical and eight kinematical
supersymmetries. In terms of the notation introduced in \cite{st1}
where a $(r,s)$-brane has $r$ Neumann directions along
$x^1,x^2,x^3,x^4$ and $s$ Neumann directions along $x^5,x^6,x^7,x^8$,
this class includes branes of type $(r,s)$ with $|r-s|=2$.
We are  here using the same conventions as in
\cite{gg}\footnote{In particular, the notation $(r,s)$ applies both to the
lorentzian signature $Dp$-branes (for which  the light-cone $x^\pm$
directions are tangential to the world-volume) and to the instantonic
brane (for which  the light-cone $x^\pm$ directions are transverse to
the world-volume).}.
\medskip

\noindent {\bf $\bullet$   Class II.} In this case the gluing matrix
satisfies
\be\label{pmpmII}
\left(\Pi M\Pi M\right)_{ab} = + \delta_{ab} \,.
\ee
The branes in this class that preserve some supersymmetry
are $(0,0)$ (the euclidean $D$-instanton or lorentzian $D$-string) and
$(0,4)$ or $(4,0)$. The latter are only consistent in the presence of
world-volume flux. These $D$-branes preserve eight dynamical
supersymmetries but do not preserve the kinematical supersymmetries.
\medskip

A separate study of $D$-branes in the context of generalised
string theory $pp$-wave backgrounds of \cite{mm} was performed in
\cite{hk} (following \cite{hiv}). In particular, backgrounds whose
world-sheet theory has (dynamical) $(2,2)$ supersymmetry were
considered, which have supersymmetric A- and B-type branes.
The B-type $D$-branes
(whose world-volume describes a holomorphic submanifold) have to
satisfy that
on the world-volume of the brane
\be
W = {\rm constant}\,,
\label{conscon}
\ee
where $W$ is the superpotential. The maximally supersymmetric
plane-wave is a special background of this type with superpotential
\be
W= i (z_1^2 + z_2^2 + z_3^2 + z_4^2)\, ,
\label{supco}
\ee
where $z_i=x^i+i x^{i+4}$. The analysis of \cite{hk} therefore implies
that any brane for which (\ref{supco}) is constant along the
world-volume preserves at least some fraction of the $(2,2)$
supersymmetry. The additional A-type branes (which are special
Lagrangian submanifolds) are class II $(4,0)$ and $(0,4)$ branes with
world-volume fluxes.

The solutions of the condition (\ref{conscon}) include {\it oblique}
$D$-branes, whose isometry group is a subgroup of the diagonal $SO(4)$
subgroup of the background symmetry. These branes cannot be
characterised in terms of the $(r,s)$-labels as above. The new
$D$-branes found in \cite{hk} are the following:
\medskip

\noindent {\it (i)  Oblique $D3$-branes.} In this case, the world-volume
is parametrised \eg by
\be\label{wv}
z_1= i z_2 \,, \quad  z_3= a\,,\quad z_4 = b \,,
\ee
where $a$ and $b$ are complex constants. The two Neumann directions
are therefore $x^1-x^6$ and $x^2+x^5$, and the corresponding gluing
matrix is
\be\label{Mmat3}
M_3 = \half (\gamma^1-\gamma^6) (\gamma^2 +\gamma^5)\,.
\ee
It is easy to see that in this case
\be\label{M3char}
\Pi M_3 \Pi M_3 = \gamma^1 \gamma^2 \gamma^5 \gamma^6 \,.
\ee
Thus $M_3$ does not satisfy either (\ref{pmpmI}) or
(\ref{pmpmII}).
\medskip

\noindent {\it (ii)  Oblique $D5$-branes.} The complex parametrisation
in this case \eg takes the form
\be\label{wv5}
z_1= i z_2 \,, \quad  z_3= iz_4  \,,
\ee
so that $c=0$ and the Neumann directions are given by
\be\label{Ndirections}
x^1-x^6\,,\ x^2+x^5\,,\ x^3-x^8\,,\ x^4+x^7\,.
\ee
The corresponding gluing matrix is
\be\label{Mmat5}
M_5 = {1\over 4} (\gamma^1 - \gamma^6) (\gamma^2 + \gamma^5)
               (\gamma^3 - \gamma^8) (\gamma^4 + \gamma^7) \,.
\ee
Again, by a direct computation one finds that
\be\label{M5char}
\Pi M_5 \Pi M_5 = \Pi \widehat{\Pi} \,.
\ee
Given our convention described above, the undotted component of
(\ref{M5char}) satisfies (\ref{pmpmI}) above, and thus the
$OD5$-brane is in fact a class I brane.
\medskip

\noindent {\it (iii)  Curved $D7$-brane.} The world-volume in this case
satisfies the constraint
\be\label{D7worldvolume}
z_1^2 + z_2^2 + z_3^2 + z_4^2 =c \,,
\ee
which describes a deformed conifold.  An analogous curved
$D5$-brane (which was not mentioned in \cite{hk}) will also be
discussed.  
\medskip

In this paper we will construct the boundary states of
the oblique $D3$-brane and oblique $D5$-brane as well as their
description in terms of open strings. We shall analyse how much
supersymmetry these branes actually preserve. We will also discuss
some aspects of the open string description of the curved
$D7$-brane although the construction of the boundary state raises new
issues which we will not address.

The guiding principle in the construction of the boundary states is
the preservation of various supersymmetries. In the cases studied
earlier, the condition that the boundary state is annihilated by half
the dynamical light-cone supercharges was imposed,
\be\label{dynamicalglue}
\left( Q_{\dot a} + i \eta\, M_{\dot a \dot b}\,
{\tilde Q}_{\dot b}\ \right)
|\!| B,\eta\,\rangle\!\rangle = 0\,.
\ee
Together with the bosonic gluing conditions, this determined the
gluing conditions for the fermionic modes. In the context of the
oblique branes, proceeding in this way can lead to inconsistent gluing
conditions on the fermions.  We will find that the oblique $D3$-brane
preserves only a subalgebra of the full supersymmetry algebra. This is
confirmed by an open string supersymmetry analysis, which will show
that only four dynamical and four kinematical supersymmetries are
preserved. In the boundary state analysis, one needs to project the
condition in equation (\ref{dynamicalglue}) onto the conserved
supersymmetries in order to get consistent conditions.  By contrast,
the oblique $D5$-brane turns out to be a class I brane which preserves
eight dynamical and eight kinematical supersymmetries. As a result the
projection on the supercharges in this case is trivial.
\medskip

The outline of this paper is as follows. In section 2, the open string
supersymmetries are analysed and the relevant projectors on the
supersymmetries are determined. The open string mode expansions are
found, and the corresponding one-loop diagrams are calculated.
In section 3, the boundary states for the oblique $D3$-brane and
$D5$-brane are constructed. Their tree-level cylinder diagrams are
determined and shown to agree with the open string one-loop diagrams.
Section 4 presents the open string description of the curved
$D7$-brane, as well as the analysis of the curved $D5$-brane. 

%%%%

\section{The open-string point of view}

Type IIB superstring theory in the maximally supersymmetric
plane-wave background \cite{met,mt} is naturally formulated in
light-cone gauge.  In this gauge the 32 supersymmetries of the
background divide into sixteen nonlinearly realised `kinematical'
supersymmetries and sixteen linearly realised `dynamical'
supersymmetries. A subset of these supersymmetries is preserved by
the oblique brane boundary conditions, as we will now see.
The light-cone gauge plane-wave IIB action is given by (using the
conventions as in \cite{gg})
\be
I=\frac{1}{2\pi \alpha'}\int
d\tau\int_{0}^{\pi}d\sigma\left[\partial_+X^I\partial_-
X^I-\hm^2(X^I)^2-(S\partial_+ S+\tS \partial_- \tS -2\hm S\Pi
\tS)\right]\,. \label{action}
\ee

The dynamical supersymmetry transformations of the coordinates are
given by
\ba\label{dyn}
\delta_{\epsilon}X^{I}&=&
\left[ \epsilon^1\g^I S+\epsilon^2 \g^I \tilde{S}\right]
\nn\\
\delta_{\epsilon}S^a&=&
\left[\partial_{-}X^I(\g^I)^{a\dot{b}}\epsilon^1_{\dot{b}}
+\hm X^I(\Pi\g^I)^{a\dot{b}}\epsilon^2_{\dot{b}}\right]  \\
\delta_{\epsilon}\tilde{S}^a&=&\left[\partial_{+}
X^I(\g^I)^{a\dot{b}}\epsilon^2_{\dot{b}}
-\hm X^I(\Pi\g^I)^{a\dot{b}}\epsilon^1_{\dot{b}}\right]
\,,\nn
\label{fermbou}
\ea
while the kinematical supersymmetry transformations are
\ba\label{kin}
\delta_{\kappa}S&=&\cos \hm\tau \,\kappa^1+\sin \hm\tau \,\Pi\kappa^2\nn\\
\delta_{\kappa}\tilde{S}&=&
\cos \hm\tau \,\kappa^2-\sin \hm\tau \, \Pi\kappa^1  \,.
\ea

The fermionic coordinates of an open string with end-points on the
brane satisfy the boundary conditions
\be
S(\sigma,\tau)|_{\sigma=0,\pi}=M
\tilde{S}(\sigma,\tau)|_{\sigma=0,\pi}\,.
\label{openbc}
\ee
The number of supersymmetries which are preserved on the $D$-brane
can be determined by the requirement that the boundary conditions
(\ref{openbc}) (as well as the Dirichlet or Neumann boundary
conditions on $X^I$) are invariant under supersymmetry
transformations,
\be
\delta S|_{\sigma=0,\pi}=M \delta \tilde{S}|_{\sigma=0,\pi} \,.
\label{susyvars}
\ee

\subsection{Dynamical supersymmetries}

If $I$ labels a Dirichlet direction, $I\in{\cal D}$, then the
supersymmetry variation of $X^I$ (\ref{dyn}) must vanish at the
boundary. Together with (\ref{openbc}) and imposing
\be
\epsilon^1 = -M \epsilon^2 \,,
\label{epsrel}
\ee
leads to the condition on the gluing matrix  $M$,
\be
M^t \gamma^I M - \gamma^I = 0\,,\qquad I\in{\cal D}\,.
\label{gluerel}
\ee
[Equivalently, we could have chosen $\epsilon_1 = M \epsilon_2$
with $M^t \gamma^I M + \gamma^I = 0$, but the end result is
the same.]

Next we require that the fermionic boundary condition (\ref{openbc})
is invariant under the supersymmetry variation. It follows directly
from (\ref{dyn}) that
\be\label{opensusyvar}
0 = \delta_\epsilon (S - M \tilde{S}) |_{\sigma=0,\pi} =
- \left[ (\partial_- X^I) \gamma^I M
+ (\partial_+ X^I) M \gamma^I \right] \epsilon^2|_{\sigma=0,\pi}
+ O(\hm) \,,
\label{epsrel2}
\ee
where $O(\hm)$ denotes terms that depend on the mass parameter, and we
have used (\ref{epsrel}). If $I$ is a Dirichlet direction,
$\partial_\tau X^I = \partial_+ X^I + \partial_- X^I=0$ at the
boundary, and the corresponding terms in the bracket of
(\ref{epsrel2}) vanish by (\ref{gluerel}). On the other hand, if $I$
labels a Neumann direction, $I\in{\cal N}$, the relevant boundary
condition is  $\partial_+ X^I -\partial_- X^I=\partial_\sigma X^I=0$,
and this leads to\footnote{In reaching this conclusion we have assumed
that the terms independent of $\hm$ have to vanish by themselves; for
branes that satisfy standard Dirichlet or Neumann boundary conditions
for the bosons, this is necessarily the case, given the structure of
the $\hm$-dependent term, see (\ref{opsus}) below. However, for example
for the case of the $(4,0)$- and $(0,4)$-brane this assumption is not
satisfied.}
\be
M^t \gamma^I M + \gamma^I = 0 \,, \qquad \qquad I\in{\cal N}\,.
\label{gluerel2}
\ee
The remaining ($\hm$-dependent) terms of (\ref{opensusyvar}) are then
\be
\delta_\epsilon (S  -M \tilde{S})|_{\sigma=0,\pi}
=
\hm\left(\sum_{J\in {\cal N}}(\Pi+M \Pi M)\gamma^J X^J
+\sum_{I\in{\cal D}}(\Pi-M \Pi M)\gamma^{I}X^{I}\right)\epsilon^2=0\,.
\label{opsus}
\ee
If the $D$-brane is located at the origin in the Dirichlet directions
the second term is zero. The condition that the first term
(associated with Neumann directions) also vanishes is
\be\label{opensusy}
\left( \bbbone + \Pi M\Pi M \right) \gamma^J \epsilon^2 =0 \,,
\label{neumcon}
\ee
for all $J\in {\cal N}$ and $\epsilon^1 = -M\epsilon^2$. For the
following it is useful to define the projectors
\be\label{projector}
P^\pm = {1\over 2} \left( 1 \pm \Pi M\Pi M \right) \,.
\ee
Consider now the case of the oblique $D3$-brane, where $M=M_3$
(\ref{Mmat3}), and (\ref{projector}) is
\be\label{projector3}
P^\pm_3 = {1\over 2}
\left(\bbbone\pm\gamma^1\gamma^2\gamma^5\gamma^6\right) \,,
\ee
as follows from (\ref{M3char}). The condition (\ref{neumcon}) can be
rewritten by commuting $\gamma^J$, for each Neumann direction $J$, to
the left of the bracket. This results in the equation
\be\label{eps2con}
P^-_3 \epsilon_2 = {1\over 2}
\left(\bbbone-\gamma^1\gamma^2\gamma^5\gamma^6\right) \epsilon^2=0\,.
\ee
Thus if $\epsilon^2$ satisfies (\ref{eps2con}), the boundary condition
is invariant under the corresponding supersymmetry transformation.

At first this conclusion only holds if the $OD3$ is located at the
origin. However, the second term in (\ref{opsus}) is
\be
\sum_{I\in{\cal D}}
X^I \Pi\,
\Bigl(\bbbone-\gamma^1\gamma^2\gamma^5\gamma^6\Bigr)\,
\gamma^I \epsilon^2
\,.
\ee
Since $\epsilon^2$ satisfies (\ref{eps2con}), it follows that the
$OD3$-brane continues to be supersymmetric if it is placed at an
arbitrary position along $x^3,x^4,x^7,x^8$. This is in agreement with
the properties of the oblique $D3$-brane in \cite{hk}, which
permitted the brane to be at arbitrary values of $z_3 = x_3+ix_7$ and
$z_4= x_4+ix_8$. It is also in agreement with the boundary state
analysis in the closed-string sector that will be considered in the
next section.

For the case of the oblique $D5$-brane, $M=M_5$ (\ref{Mmat5}), and
$\Pi M_5\Pi M_5=\Pi\widehat\Pi$ as was already observed in
(\ref{M5char}). Since $\epsilon^2$ has a dotted index and $\gamma^J$
changes the chirality of the spinor, the first bracket in equation
(\ref{opensusy}) has undotted indices. In our conventions,
$\Pi\widehat\Pi=-\bbbone$ for the undotted indices, and thus
(\ref{opensusy}) is satisfied without any further constraint on
$\epsilon_2$. It is also easy to see that the brane is confined to the
origin (again in agreement with \cite{hk}). Both of these
statements actually follow from the fact that the gluing matrix
$M_5$ satisfies the class I condition (\ref{pmpmI}) for the undotted
indices.

\subsubsection{A more general analysis}

To analyse in general which gluing matrices $M$ will preserve a
certain fraction of the dynamical supersymmetries, it is convenient to
introduce, for each Neumann direction $X^J$, the matrix
\be\label{Kdef}
K^J = \gamma^J \Pi M \Pi M \gamma^J \,,
\ee
so that $\Pi M\Pi M \gamma^J = \gamma^J K^I$.
With this notation (\ref{opensusy}) becomes
\be\label{susyfrac}
{1\over 2} \left( \bbbone + K^J \right)  \epsilon_2 =0 \,.
\ee
A necessary condition for this to have non-trivial solutions (except
for the case, when there are no Neumann directions, as is the case for
the $(0,0)$-brane) is that
\be\label{susycond}
\bbbone = K^J K^J = \gamma^J K^J K^J \gamma^J =
\left(\Pi M\Pi M \right)^2 \,.
\ee
The fraction of dynamical supersymmetry that is preserved is
determined by the dimension of the joint kernel of the different
projectors in (\ref{susyfrac}).

Since $M$ is a product of an even number of gamma matrices, it follows
from (\ref{Kdef}) that $K^I$ can be written as a sum of products of
$0$, $4$ or $8$ gamma matrices. For the usual class I branes,
$K^J=-\bbbone$ for all $J$, and the kernel of the projectors has
dimension 8 trivially. On the other hand, for the oblique $D5$-brane,
each $K^J=\Pi\widehat{\Pi}$, and thus again the kernel of the
projectors has dimension $8$. The situation is different for the class
II branes where $K^J=\bbbone$ for all $J$. 
%A
The only supersymmetric
$D$-branes are then the $D$-instanton (which does not have any Neumann
directions) and the $(4,0)$ and $(0,4)$ branes, for which the above
analysis does not apply because of their world-volume flux.
%E
In case of the (genuinely) oblique
branes $K^J$ is a product of 4 gamma matrices and when $K^J$ is the same
for all Neumann directions (as is the case for the $OD3$-brane), a
quarter of the dynamical supersymmetries is preserved.

\subsection{Kinematical supersymmetries}

Equations (\ref{kin}), together with the boundary conditions, lead to
\be
\kappa^1=M \kappa^2, \qquad
P^+ \kappa^2 = {1\over 2}(\bbbone+\Pi M \Pi M)\kappa^2=0\, .
\label{kinsymm}
\ee
Thus some kinematical supersymmetries are preserved if the dimension
of the kernel of ${1\over 2} (\bbbone+\Pi M \Pi M)$ is non-zero.
For the $OD3$ case
$\Pi M_3 \Pi M_3=\gamma^1 \gamma^2 \gamma^5\gamma^6$, and the
dimension of the kernel of $P^+_3$ is four. This implies that the
$OD3$-brane preserves four kinematical supersymmetries, which is one
quarter of the background kinematical supersymmetries.
For the $OD5$ case $\Pi M_5 \Pi M_5=\Pi \widehat{\Pi}$, so that
${1\over 2}(\bbbone+\Pi M_5 \Pi M_5)$ vanishes trivially (since we
have made the chirality choice  $\Pi \widehat{\Pi}=-\bbbone$ for
undotted spinors, and (\ref{kinsymm}) is again an undotted
equation). Therefore, in  this case eight kinematical supersymmetries
are preserved, as usual for a class I brane.

\subsection{Open string mode expansion for the $OD3$-brane}
\label{openloop}

In order to be able to determine the various cylinder diagrams it is
useful to describe the boundary condition (\ref{openbc}) in terms of
modes. (In the following we shall only consider the fermionic fields;
the mode expansion for the bosonic fields are as usual. ) As is
explained in \cite{gg}, the fermionic fields have the mode expansion
\begin{eqnarray}\label{modeexpansion}
S(\sigma,\tau) & = & S_0' \cos(\hm\tau)
+ \Pi\tilde{S}_0' \sin(\hm\tau)
+ S_0 \cosh(\hm\sigma) + \Pi \tilde{S}_0 \sinh(\hm\sigma)
\nn\\
& & \qquad + \sum_{n\ne 0} c_n \left[
S_n\,  e^{-i(\hom_n\tau - n \sigma)}
+   {i\over \hm} (\hom_n-n) \Pi \tilde{S}_n
        e^{-i(\hom_n\tau + n \sigma)} \right] \,, \\
\tilde{S}(\sigma,\tau) & = &
-\Pi S_0' \sin(\hm\tau) + \tilde{S}_0' \cos(\hm\tau)
+ \tilde{S}_0 \cosh(\hm\sigma) + \Pi S_0 \sinh(\hm\sigma)  \nn\\
& & \qquad + \sum_{n\ne 0} c_n \left[
    \tilde{S}_n e^{-i(\hom_n\tau + n \sigma)}
- {i\over \hm} (\hom_n-n) \Pi S_n e^{-i(\hom_n\tau - n \sigma)}
\right]\,,
\end{eqnarray}
where $\hom_n=\sqrt{\hm^2+n^2}$, and $\hm$ is the mass parameter in
the associated open string light cone gauge.

Let us first analyse the open string both of whose endpoints lie on an
oblique $D3$-brane. For the zero modes (\ref{openbc}) with $M=M_3$
(\ref{Mmat3}) implies
\ba\label{Sobc}
&& S_0' = M_3 \tilde{S}_0' \,,\qquad
\tilde{S}_0' = - \Pi M_3 \Pi S_0' \,,\\
&& S_0 = M_3  \tilde{S}_0 \,,\qquad
\tilde{S}_0 = \Pi M_3 \Pi S_0 \,.
\ea
Given (\ref{M3char}), these equations are only consistent provided
we project $S_0'$ and $S_0$ onto the components $P_3^- S_0'$, and
$P_3^+ S_0$, where $P^\pm_3$ was defined in (\ref{projector3}). Thus
this open string has four zero modes $P_3^+ S_0$, and four zero modes
$P^-_3S'_0$.

For the non-zero modes the boundary condition (\ref{openbc}) with
$M=M_3$ implies
\be\label{Snbc}
S_n = {\cal N}_n \tilde{S}_n \,,
\ee
where $n\in \Zop$ and
\be\label{calN}
{\cal N}_n =
\left( \bbbone + i \,{(\hom_n - n)\over \hm}\, M_3 \Pi \right)^{-1}
\left( \bbbone - i \,{(\hom_n - n)\over \hm}\, \Pi M_3^t\right)
M_3 \,.
\ee
By using that $\Pi^\dagger\Pi=\bbbone$ and $M_3^\dagger= M_3^t=-M_3$,
it  follows directly that ${\cal N}_n$ is a unitary matrix,
\ba
{\cal N}_n {\cal N}_n^\dagger
&=& \bbbone \,.
\ea
In order to write ${\cal N}_n$ more explicitly, it is convenient to
write (\ref{Snbc}) separately for the $P_3^\pm$ components. Since
$P^\pm_3$ commutes with $M_3$ and $\Pi$,
\be\label{comrels}
[P^\pm_3,M_3 ] = [P^\pm_3,\Pi] = 0 \,,
\ee
(\ref{Snbc}) can be written
as
\be\label{Snbc1}
P^\pm_3 S_n = {\cal N}_n^\pm\;  P^\pm_3 \tilde{S}_n \,,
\ee
where
\be
{\cal N}_n^+ = {1\over \hom_n} \left[ n \bbbone - i \hm M_3 \Pi
\right] M_3 \,, \qquad
{\cal N}_n^- = M_3 \,.
\ee
Thus it follows that the minus components behave just as for a usual
class I brane, while the behaviour of the plus components is as for a
class II brane. Because of the four fermionic zero modes $P_3^+S_0$,
the one loop (cylinder) diagram of this open string vanishes.
\medskip

For a string stretching between a $OD3$-brane and its anti-brane
$\overline{OD3}$ none of the zero modes above survives. As regards the
non-zero modes, the conditions arising at $\sigma=0$ are
(\ref{Snbc1}), while at $\sigma=\pi$, one finds
\be\label{barSnbc}
P_3^\pm S_n = - e^{2\pi i n } \,\widehat{\cal N}_n^\pm \,
P_3^\pm \tilde{S}_n \,,
\ee
where now
\be
\widehat{\cal N}_n^+ =
{1\over \hom_n} \left[ n \bbbone + i \hm M_3 \Pi
\right] M_3 \,, \qquad
\widehat{\cal N}_n^- = M_3 \,.
\ee
In order to solve (\ref{Snbc1}) and (\ref{barSnbc}) simultaneously,
the minus components $P_3^- S_n$ must have $n\in \Zop+\half$, while
two of the plus components $P_3^+ S_n$ satisfy
\be\label{plus1}
{n - i \hm \over n + i \hm} = - e^{2\pi i n } \,,
\ee
while the other two plus components satisfy
\be\label{plus2}
{n + i \hm \over n - i \hm} = - e^{2\pi i n } \,.
\ee
In fact, the two components that satisfy (\ref{plus1}) are
the two eigenstates with eigenvalue $+1$ under the action of $\Pi$,
while for the two components in (\ref{plus2}) the eigenvalue of
$\Pi$ is $-1$. The solution $n=0$ is to be discarded in either
case. As in \cite{gg} one can thus deduce that the open string
one-loop diagram in this case equals
\be\label{openanti}
{\cal A}_{\overline{OD3},OD3} =
{ \left(\hat{g}_4^{(\hm)}(\tilde{q}) \right)^2
\left( f_2^{(\hm)}(\tilde{q}) \right)^4 \over
\left( f_1^{(\hm)}(\tilde{q})\right)^8 } \,,
\ee
where the functions $f_i^{(\hm)}$ and $\hat{g}_4^{(\hm)}$ are those
defined in \cite{bgg,gg}, and we have assumed, for simplicity, that
both branes are located at the origin in the transverse space.

%%%%

\section{Boundary states for the  oblique branes}

We now turn to the description of these oblique $D$-branes in
terms of boundary states.  These are closed-string states that
describe the coupling of the brane to arbitrary closed-string states.
This is most simply formulated in the light-cone gauge with the
$X^\pm$ directions transverse to the brane \cite{ggut} --- in other
words, for branes with euclidean world-volumes, or instantonic branes.
As argued in \cite{bgg}, the description of the supersymmetry
conditions for a euclidean brane with $(p-1)$-dimensional world-volume
is closely related to that of a lorentz signature $D(p+1)$-brane with
longitudinal $X^\pm$. The dynamical supercharges (for which suitable
superpositions are preserved by the boundary) are explicitly given
as (see \cite{gg} for an explanation of the notation)
\ba\label{Qff}
\sqrt{ 2 p^+}\, Q_{\dot a}
&=&  p_0^I \gamma^I_{\dot a b} S_0^b
   - m x_0^I \left(\gamma^I\Pi\right)_{\dot a b}
  {\tilde S}_0^b
\\
&+& \sum_{n=1}^\infty
\left( c_n \gamma^{I}_{\dot a b}
       (\alpha_{-n}^I S_n^b + \alpha_n^I S^{b}_{-n} )
+
 \frac{{i} m}{2\omega_n c_n }
 \left(\gamma^I\Pi\right)_{\dot a b}
       ({\tilde \alpha}_{-n}^I {\tilde S}_n^b
        -{\tilde \alpha}_n^I {\tilde S}^b_{-n} )
\right)\,,
\nonumber\\
\label{Qbarff}
\sqrt{ 2 p^+}\,  {\tilde Q}_{\dot a}
&=& p_0^I \gamma^{I}_{\dot a b} {\tilde S}_0^b
  + m x_0^I\left(\gamma^I\Pi\right)_{\dot a b}{ S}_0^b
\\
&+& \sum_{n=1}^\infty
\left(( c_n  \gamma^{I}_{\dot a b}
       ({\tilde \alpha}_{-n}^I {\tilde S}_n^b
        +{\tilde \alpha}_n^I {\tilde S}^b_{-n} )
-
 \frac{{i} m}{2 \omega_n c_n }
 \left(\gamma^I\Pi\right)_{\dot a b}
       (\alpha_{-n}^I S_n^b - \alpha_n^I S^b_{-n})
\right)\,.
\nonumber
\end{eqnarray}
They satisfy the superalgebra
\be\label{Qalg}
\{ Q^+_{\dot a}, Q^-_{\dot b} \}  = 2\, \delta_{\dot a\dot b}\, H +
m\, (\gamma^{ij} \,\Pi)_{\dot a \dot b} \, J^{ij}
+ m \, (\gamma^{i'j'}\, \Pi)_{\dot a \dot b} \, J^{i'j'}  \,,
\ee
where
$Q^\pm_{\dot a}={1\over\sqrt{2}}(Q_{\dot a} \pm i \tilde{Q}_{\dot a})$.

\subsection{The $OD3$-brane}

Let us begin by describing the oblique $D3$-brane. For definiteness,
we consider the brane that has Neumann directions along $x^1-x^6$ and
$x^2+x^5$, whose gluing matrix is given by $M_3$ in (\ref{Mmat3}). Any
other $OD3$-brane can be obtained from this by a
$SO(4)\times SO(4)$ rotation.

The gluing conditions for the bosonic modes are obtained in a standard
fashion as
\ba\label{bosonicglue}
\left( a^i_n - \tilde{a}^i_{-n} \right) \,
|\!| D,\eta\,\rangle\!\rangle &=& 0 \,,\quad i\in {\cal D} \nn \\
\left( a^i_n + \tilde{a}^i_{-n} \right) \,
|\!| D,\eta\,\rangle\!\rangle &=& 0
\,,\quad i\in{\cal N}\,.
\ea
Here $a^i_n$ are the modes appropriate to an adapted orthogonal basis;
for example, for the Neumann direction $i=1-6$,
$a^i_n = \alpha^1_n-\alpha^6_n$, where $\alpha_n^i$ are the modes of
the bosonic field $X^i(z)$.

Given the analysis of the open string, the boundary states we are
interested in should satisfy a projected supercharge condition of the
form
\be\label{susycondition}
P_3^+ \left(Q + i \eta\, M_3\, {\tilde Q} \right)
|\!| OD3,\eta\,\rangle\!\rangle = 0\,.
\ee
The strategy of the construction (as in \cite{gg}) is to determine the
fermionic gluing conditions from the supercharge condition and the
bosonic gluing conditions.

\subsubsection{The zero mode gluing conditions}

We will first consider the zero mode part of the condition
(\ref{susycondition}), \ie\ the terms proportional to $p_0^I$ and
$x_0^I$. If the brane is located at the origin in the transverse
directions, the conditions for the Dirichlet directions (\ie\ the
terms proportional to $p_0^I$, where $I$ is a Dirichlet direction) are
\be\label{closedzeromodeglue}
P_3^+ \left(\gamma^I  \, S_0 + i\eta\, M_3\, \gamma^I \,
\tilde{S}_0\right)
\, |\!| OD3,\eta\,\rangle\!\rangle = 0\,,
\ee
where $I$ is a Dirichlet direction. Since
$\gamma^I M_3 = M_3 \gamma^I$, (\ref{closedzeromodeglue}) simplifies
to
\be\label{closedzeromodeglue1}
P_3^+ \gamma^I \left(S_0 + i\eta \, M_3 \, \tilde{S}_0\right)
\, |\!| OD3,\eta\,\rangle\!\rangle = 0\,.
\ee
If $I$ is a Dirichlet direction other than
$x^1+x^6$ or $x^2-x^5$ (\ie\ $I=3,4,7,8$), then $\gamma^I$ commutes
with $P_3^+$. Multiplying (\ref{closedzeromodeglue1}) by $\gamma^I$
then implies
\be\label{closedzeromodeglue2}
P_3^+ \left(S_0 + i\eta\, M_3 \, \tilde{S}_0\right)
\, |\!| OD3,\eta\,\rangle\!\rangle = 0\,.
\ee
On the other hand, if $I$ corresponds to the Dirichlet
directions $x^1+x^6$ or $x^2-x^5$, then
$P_3^+ \gamma^I  = \gamma^I P_3^-$. Multiplying
(\ref{closedzeromodeglue1}) by $\gamma^I$ then leads to
\be\label{closedzeromodeglue3}
P_3^- \left(S_0 + i\eta\, M_3 \, \tilde{S}_0\right)
\, |\!| OD3,\eta\,\rangle\!\rangle = 0\,.
\ee
Adding this to (\ref{closedzeromodeglue2}) it  follows that the
zero modes must satisfy the condition
\be\label{closedzeromodeglue4}
\left(\,S_0+i\eta\, M_3 \tilde{S}_0 \,\right)\,
|\!| OD3,\eta\,\rangle\!\rangle =0 \,.
\ee

We must now check whether (\ref{closedzeromodeglue4}) is compatible
with the constraints that arise for the Neumann directions, for which
$p^J_0=0$ on the boundary state. The zero mode part of
(\ref{susycondition}) (\ie\ the constraints coming from the terms
proportional to $x_0^J$ with $J$ a Neumann direction) implies
\be\label{neucond}
P_3^+ \left( M_3\, \gamma^J \, \Pi\,  S_0
+ i \eta\, \gamma^J\, \Pi\, \tilde{S}_0 \right)
\, |\!| OD3,\eta\,\rangle\!\rangle =0 \,.
\ee
Using $M_3\gamma^J = - \gamma^J M_3$ this can be rewritten as
\be\label{neucond1}
- P_3^+\, \gamma^J\, M_3\, \Pi \left(
S_0 - i \eta \,\Pi M_3^t \Pi \, \tilde{S}_0 \right) \,
|\!| OD3,\eta\,\rangle\!\rangle = 0 \,.
\ee
Using the fact that $P_3^+ \gamma^J = \gamma^J P_3^-$ for Neumann
directions, this condition is therefore equivalent to
\be\label{neucond2}
P_3^- \left(
S_0 - i \eta \, \Pi M_3^t \Pi\, \tilde{S}_0 \right) \,
|\!| OD3,\eta\,\rangle\!\rangle = 0 \,,
\ee
which is compatible with (\ref{closedzeromodeglue3}) since
\be\label{minusid}
P_3^- M_3 = - P_3^- \Pi M_3^t \Pi\,.
\ee

Finally, we  shall analyse whether the condition $x^I=0$ when $I$ is a
Dirichlet direction can be relaxed. The term that is proportional to
$x^I$ in (\ref{susycondition}) is
\be\label{position}
P_3^+ \left( M_3\, \gamma^I\, \Pi\, S_0
+ i \eta\,  \gamma^I \, \Pi\, \tilde{S}_0 \right)
\, |\!| OD3,\eta\,\rangle\!\rangle =0 \, .
\ee
This is the same as (\ref{neucond}) but now $\gamma^I$ commutes with
$M_3$, and therefore (\ref{position}) becomes
\be\label{position1}
P_3^+ \,\gamma^I\, M_3\, \Pi \left(S_0 + i \eta\,  \Pi M_3^t \Pi\,
\tilde{S}_0 \right)\,
|\!| OD3,\eta\,\rangle\!\rangle =0 \,.
\ee
As before there are two cases to distinguish: if the Dirichlet
direction corresponds to $I=3,4,7,8$ then $\gamma^I$ commutes with
$P_3^+$. Multiplying (\ref{position1}) by $\gamma^I$ as well as
$\Pi M_3^{t}$ then implies
\be\label{position2}
P_3^+ \left(S_0 + i \eta \,\Pi M_3^t \Pi\,  \tilde{S}_0 \right)\,
|\!| OD3,\eta\,\rangle\!\rangle =0 \,.
\ee
This is compatible with (\ref{closedzeromodeglue2}) since
\be
P_3^+ M_3 = + P_3^+ \Pi M_3^t \Pi\,.
\ee
On the other hand, if $I$ is either $x^1+x^6$ or $x^2-x^5$, then
(\ref{position1}) becomes
\be\label{position3}
P_3^- \left(S_0 + i \eta\,\Pi M_3^t \Pi\, \tilde{S}_0 \right)\,
|\!| OD3,\eta\,\rangle\!\rangle =0 \,,
\ee
which is incompatible with (\ref{closedzeromodeglue3}) because of
(\ref{minusid}). It therefore follows that the oblique brane can be
moved away from the origin along any of  the $I=3,4,7,8$ directions
(which are transverse to the $1-2-5-6$ plane) but that they are stuck
at $x^1=-x^6$ and $x^2=x^5$.\footnote{Using the isometry of the
background, one can obviously move these branes away from the
origin. However, the resulting branes then become explicitly
time-dependent \cite{st2}.} This is in agreement with the open string
analysis mentioned earlier, as well as the results of \cite{hk}.

In order to characterise the fermionic ground state of the boundary
state, it is convenient to rewrite (\ref{closedzeromodeglue4})
separately for the $P_3^+$ and $P_3^-$ components. Using
\be
P_3^+ M_3 = P_3^+ \gamma^1 \gamma^5 \,, \qquad
P_3^- M_3 = P_3^- \gamma^1 \gamma^2 \,,
\ee
it follows that
\be\label{zerop}
P_3^+ \left(\,S_0+i\eta \,\gamma^1 \gamma^5\, \tilde{S}_0 \,\right)\,
|\!| OD3,\eta\,\rangle\!\rangle =0 \,,
\ee
and
\be\label{zerom}
P_3^- \left(\,S_0+i\eta\, \gamma^1 \gamma^2 \, \tilde{S}_0 \,\right)\,
|\!| OD3,\eta\,\rangle\!\rangle =0 \,.
\ee

\subsubsection{Gluing conditions for the non-zero modes}

For each Dirichlet direction (labeled by $I$) the gluing conditions
that follow from (\ref{susycondition}) after substituting
$\tilde \alpha_n^I = \alpha_{-n}^I$ (from (\ref{bosonicglue}))
and equating coefficients of $\alpha_n^I$ are
\be\label{Dnz1}
P_3^+ \gamma^I \left[
\left( \bbbone + {m\eta \over 2 \omega_n c_n^2} M_3 \Pi \right) S_n
+ i \eta
\left( \bbbone - {m\eta \over 2 \omega_n c_n^2} \Pi M_3^{t} \right)
M_3 \, \tilde{S}_{-n} \right] \,
|\!| OD3,\eta\,\rangle\!\rangle =0 \,.
\ee
For each Neumann direction (labeled by $J$) the gluing conditions that
follow from (\ref{susycondition}) after substituting
$\tilde \alpha_n^I = - \alpha_{-n}^I$ are
\be\label{Nnz1}
P_3^+ \gamma^J \left[
\left( \bbbone - {m\eta \over 2 \omega_n c_n^2} M_3 \Pi \right) S_n
+ i \eta
\left( \bbbone + {m\eta \over 2 \omega_n c_n^2} \Pi M_3^{t} \right)
M_3\, \tilde{S}_{-n} \right] \,
|\!| OD3,\eta\,\rangle\!\rangle =0 \,.
\ee
By considering the two types of Dirichlet directions (\ie\
$I=3,4,7,8$ or $I=1+6,2-5$), the same arguments as in the
previous subsection show that (\ref{Dnz1}) is equivalent to
\be\label{Dnz2}
\left[
\left( \bbbone + {m\eta \over 2 \omega_n c_n^2} M_3 \Pi \right) S_n
+ i \eta
\left( \bbbone - {m\eta \over 2 \omega_n c_n^2} \Pi M_3^{t} \right)
M_3 \, \tilde{S}_{-n} \right] \,
|\!| OD3,\eta\,\rangle\!\rangle =0 \,.
\ee
On the other hand, as with the zero modes, the Neumann condition
(\ref{Nnz1}) is equivalent to
\be\label{Nnz2}
P_3^- \left[
\left( \bbbone - {m\eta \over 2 \omega_n c_n^2} M_3 \Pi \right) S_n
+ i \eta
\left( \bbbone + {m\eta \over 2 \omega_n c_n^2} \Pi M_3^{t} \right)
M_3 \, \tilde{S}_{-n} \right] \,
|\!| OD3,\eta\,\rangle\!\rangle =0 \,.
\ee
Using (\ref{minusid}) it follows that
\be\label{DNnz}
P_3^- \left(S_n + i \eta\, M_3\, \tilde{S}_{-n} \right) \,
|\!| OD3,\eta\,\rangle\!\rangle =
P_3^- \left(S_n + i \eta\, \gamma^1 \gamma^2\, \tilde{S}_{-n} \right) \,
|\!| OD3,\eta\,\rangle\!\rangle =
0 \,.
\ee
Thus the four components of $P_3^-S_n$ are related to
$P_3^- \tilde S_{-n}$ by conventional gluing conditions of the same
type as those of a class I $(2,0)$-brane.

As regards the $P_3^+$ components, the $P_3^+$ projection of
(\ref{Dnz2}) can be rewritten as
\be\label{Dnz4}
P_3^+ \left[ S_n
+ i \eta {1\over n}
\left(\omega_n  \bbbone - m\eta\, M_3 \Pi \right) M_3 \,
\tilde{S}_{-n} \right] \,
|\!| OD3,\eta\,\rangle\!\rangle =0 \,.
\ee
This relates the four components of $P^+ S_n$ to $P^+ \tilde S_{-n}$
as for a class II brane \cite{gg}. Finally, it may be worthwhile
pointing out that these gluing conditions can be summarised in terms
of the fields as
\be\label{gluefields}
\left. \left( S(\sigma,\tau) + i \eta\, M_3\, \tilde{S}(\sigma,\tau)
\right) \right|_{\tau=0}  \,
|\!| OD3,\eta\,\rangle\!\rangle =0 \,.
\ee

\subsubsection{The cylinder diagrams involving the $OD3$-brane}

In section~\ref{openloop} the one-loop diagrams for the open string
stretching between two $OD3$-branes, and the open string between an
$OD3$-brane and its anti-brane were calculated. In particular, it was
found that the former diagram vanished, while the latter was given by
(\ref{openanti}). We want to show now that these diagrams are
correctly reproduced by the above boundary states, using the open
closed duality relation. (See \cite{bgg,gg} for a detailed explanation
of this relation.)

The contribution coming from the non-zero modes of the boundary states
is straightforward, but the analysis of the zero-modes requires some
care. Since the gluing conditions do not mix the plus and minus
components\footnote{Here `plus' and `minus' refers to the eigenvalue
under the action of $\Pi M_3 \Pi M_3$.}
the total contribution coming from the zero-modes is
simply the product of the two contributions from plus- and
minus-components, respectively. The $OD3$-brane satisfies the
class I gluing condition (\ref{zerom}) for the minus-components, and
their contribution is therefore $(1\pm q^m)^2$, where the sign depends
on whether we consider the overlap between the brane and the brane, or
the brane and the anti-brane \cite{bgg}. In order to understand the
contribution from the plus components, we rewrite the condition
(\ref{zerop}) in terms of the $\theta$ and $\bar\theta$ modes (as in
\cite{gg}) as 
\be\label{zerop1}
P_3^+ \left(\,\theta - \eta \,\gamma^1 \gamma^5\, \bar\theta \,\right)\,
|\!| OD3,\eta\,\rangle\!\rangle^{(0)} =0 \,,
\ee
where $|\!| OD3,\eta\,\rangle\!\rangle^{(0)}$ denotes the zero-mode
component of the boundary state. Since $\gamma^1 \gamma^5$
anticommutes with $\Pi$, (\ref{zerop}) relates $\theta_L$ to
$\bar\theta_R$, and vice versa,
\be\label{zerop2}
\begin{array}{rcl}
P_3^+ \left(\,\theta_L - \eta \,\gamma^1 \gamma^5\, \bar\theta_R
\,\right)\,
|\!| OD3,\eta\,\rangle\!\rangle^{(0)} & = & 0 \\
P_3^+ \left(\,\theta_R - \eta \,\gamma^1 \gamma^5\, \bar\theta_L \,
\right)\, |\!| OD3,\eta\,\rangle\!\rangle^{(0)} & = & 0 \,.
\end{array}
\ee
The operators that appear in the first line are both annihilation
operators, while those in the second line are both creation
operators. The boundary state is therefore created from the ground
state of the Fock space by the action of the creation operators
of the second line. Since there are two such operators, each of which
raises the energy by $m/2$, the resulting state has energy $m$. Since
this state has definite eigenvalue under the closed string
hamiltonian, the overlap between two brane states (\ie\ two states
with the same value for $\eta$) vanishes, while the overlap between
the brane state and its anti-brane state is proportional to $q^m$.
Together with the contribution from the non-zero modes, the whole
cylinder amplitude between brane and brane vanishes, while the overlap
between a brane and its anti-brane is
\be\label{antiselfoverlap}
\tilde{{\cal A}}_{\bar{D}3_o, D3_o}= \,
{\left( g_2^{(m)}(q) \right)^2 \left( f_2^{(m)}(q) \right)^4 \over
\left( f_1^{(m)}(q) \right)^8 }\,,
\ee
where $g_2^{(m)}$ is the function defined in \cite{gg}. Under the
modular transformation $t\mapsto \tilde{t}=1/t$, this reproduces
indeed (\ref{openanti}).

\subsection{The $OD5$-brane}

Next let us consider the $OD5$-brane with Neumann directions
\be\label{Ndirections5}
x^1-x^6\,,\ x^2+x^5\,,\ x^3-x^8\,,\ x^4+x^7\,,
\ee
whose gluing matrix $M_5$ was already given in (\ref{Mmat5}).
A general oblique $D5$-brane can be obtained from the above by rotations
in $SO(4)\times SO(4)$.

As we have discussed before, the $OD5$-brane behaves in fact
as a class I brane, and thus the construction of the boundary states
is as described in \cite{bgg,bp}. In particular, the gluing conditions
for the fermions are simply
\be
\left(S_n + i \eta M_5 \tilde{S}_{-n} \right) \,
|\!| OD5,\eta\,\rangle\!\rangle =0 \,,
\ee
and it is easy to see that, together with the gluing conditions for
the bosons, this leads to
\be
\left(Q + i \eta M_5 \tilde{Q} \right) \,
|\!| OD5,\eta\,\rangle\!\rangle =0 \,.
\ee
Given that the oblique $D5$-brane is in fact a class I brane, one may
wonder whether it is in fact related by a rotation to one of the
standard class I branes discussed before; this will be discussed in
the next subsection.
%%%%

\subsubsection{Rotation symmetries}

The maximally supersymmetric background is invariant under the
$SO(4)\times SO(4)$ subgroup of the transverse $SO(8)$ rotation
group. The image of any supersymmetric brane under a rotation $R$ in
$SO(4)\times SO(4)$ therefore describes another supersymmetric
$D$-brane of the same kind. Indeed, under any rotation $R$, the brane
characterised by $M$ is mapped to a brane described by
$\widehat{M} = R^t M R$. If $R$ is an element in $SO(4)\times SO(4)$,
$R$ commutes with $\Pi$, and hence
\be
\Pi \widehat{M} \Pi \widehat{M} = \Pi\, R^t M R \,\Pi\, R^t M R
                                = R^t\, \Pi M \Pi M \, R \,.
\ee
Thus if $M$ is of class I, so is $\widehat{M}$, and similarly for
class II and oblique branes.

While elements in $SO(4)\times SO(4)$ map branes of class I into
themselves, they are not the only transformations with this
property. Suppose $R$ is an element in $SO(8)$ that is not in
$SO(4)\times SO(4)$. By composing $R$ with elements in
$SO(4)\times SO(4)$, we may, without loss of generality assume, that
$R$ has the property that
\be
\Pi R = R^t \Pi \,.
\ee
The elements of $SO(8)$ with this property define the coset
$SO(8) / (SO(4)\times SO(4))$.

The condition that $\widehat{M}=R^t M R$ defines again a class I brane
(provided that $M$ does) then yields the constraint
\be\label{general}
(R^t)^2 M R^2=M\,.
\ee
Since rotations that preserve the world-volume of a brane do not map
different branes into one another, we may furthermore restrict
attention to those rotations that are of the form
$R= \prod \exp(\frac{\theta_{ij}}{2} \gamma^i \gamma^j)$ where
$i\in {\cal N}$ and $j \in {\cal D}$. Then $M R = R^t M$, and
(\ref{general}) becomes
\be
R^4=\bbbone \,. \label{Ucond}
\ee
Since the class I condition (\ref{pmpmI}) only refers to the undotted
indices, (\ref{Ucond}) only has to hold for them.

For a given brane of class I, there are only finitely many rotations
that satisfy these constraints. For example, we can take $R$ to be of
the form
\be
R=\exp\left(\frac{\theta_1}{2} \gamma^{i_1}\gamma^{j_1}\right)
  \exp\left(\frac{\theta_2}{2} \gamma^{i_2}\gamma^{j_2}\right)\,,
\label{genU}
\ee
where $i_1\ne i_2\in {\cal N}$ and $j_1\ne j_2 \in {\cal D}$. The
condition (\ref{Ucond}) is then obviously satisfied if
$\theta_1=\theta_2=\pi$; this corresponds to a parity transformation.
However, (\ref{Ucond}) is also satisfied if
$\theta_1=\theta_2=\pi/2$. The resulting transformation then maps a
class I $(r,s)$-brane into a class I $(s,r)$-brane.

Let us now return to the question of whether the oblique $D5$-brane
can be obtained from a class I brane by a rotation. Consider the class
I $(3,1)$-brane with $M=\gamma^1\gamma^2\gamma^3\gamma^7$, and define
the rotation group element $R$ by
\be
R= R^{(1,6,-)} R^{(2,5,+)} R^{(3,8,-)} R^{(4,7,-)} \,,
\ee
where for $m<n$
\be\label{rot}
R^{(m,n,\pm)}
= {1\over \sqrt{4+2\sqrt{2}}}
    \left( (1+\sqrt{2}) \bbbone \pm \gamma^m \gamma^n \right)
= \cos ({\theta/2}) \, \bbbone \pm  \sin({\theta/2})\,
\gamma^1\gamma^6 \,,
\ee
with $\theta=\pi/4$. It is easy to check that
\be
R^t \gamma^1 R = {1\over \sqrt{2}} (\gamma^1-\gamma^6) \,,
\ee
and
\be
R^t \gamma^6 R = {1\over \sqrt{2}} (\gamma^1+\gamma^6)\,.
\ee
Similarly one analyses the other directions. Overall, this rotation
therefore transforms $M$ into
\be
M_5 = R^t M R \,.
\ee
$R$ is of the form discussed before and satisfies
\be
R^4=-\Pi\widehat{\Pi}\,.
\ee
This equation reduces to $R^4=\bbbone$ for the undotted indices, and
thus satisfies the above conditions.

It follows from (\ref{rot}) that $R$ describes a rotation by $\pi/4$
in each of the four planes $x^1-x^6$, $x^2-x^5$, $x^3-x^8$ and
$x^4-x^7$. In particular, $R$ therefore maps the $(3,1)$-brane
extended along $x^1,x^2,x^3,x^7$ to the oblique $D5$-brane whose
world-volume extends along the directions (\ref{Ndirections5}).

\subsection{The projected superalgebra}

The different supersymmetric $D$-branes that we have considered above
preserve linear combinations of supercharges of the form
$(Q+iM\tilde{Q})$, or for the case of the $OD3$-brane, some
projection of this combination (\ref{susycondition}). One may thus ask
which subalgebra of the supersymmetry algebra is generated by these
generators. If we define $q= Q+iM\tilde{Q}$, one finds
\ba
\{q_{\dot{a}},q_{\dot{b}}\}
&=&2(N-\tilde{N})\delta_{\dot{a}\dot{b}} \nonumber \\
&& -\left[\frac{m}{2}
\left((\gamma^{ij}\Pi M^t)_{\dot{a}\dot{b}}J^{ij}+
(\gamma^{i'j'}\Pi M^t)_{\dot{a}\dot{b}}J^{i'j'}\right)
+(\dot{a}\leftrightarrow\dot{b})\right]\,, \label{projalg}
\ea
where $i,j=1,\ldots, 4$, $i',j'=5,\ldots, 8$, and $N, \tilde{N}$ are
the left and right-moving number operators; their difference always
vanishes on the boundary state since it imposes the level-matching
condition.\footnote{Terms such as $N-\tilde{N}$ that vanish on
physical states have been ignored in \cite{mt}.} For class I branes,
the rotation generators that appear in the second line involve
rotations that either act on the world-volume directions of the
corresponding brane, or on the space transverse to the world-volume. In
either case, this is a symmetry of the $D$-brane since the class I
branes are required to be at the origin in the transverse space. For
example, for the case of the class I $(2,0)$-brane with
$M=\gamma^1\gamma^2$, one finds
\be
\{q_{\dot{a}},q_{\dot{b}}\}=2(N-\tilde{N}) \delta_{\dot{a}\dot{b}}
+m \left[ \delta_{\dot{a}\dot{b}} J^{34} - \Pi_{\dot{a}\dot{b}} J^{12}
- \left(\gamma^{i' j'} \gamma^3 \gamma^4\right)_{\dot{a}\dot{b}}
J^{i'j'} \right] \,.
\ee
On the other hand, for the class II $D$-instanton, the terms in the
second line of (\ref{projalg}) vanish; this is as expected, since the 
$D$-instanton can be moved to any position in the transverse space,
and is therefore, in general, not invariant under any rotation.

For the case of the oblique $D5$-brane described by the matrix $M$
with
\be
M = {1\over 4} \prod_{i=1}^4 (\gamma^i \pm \gamma^{\hat{\imath}} ) \,,
\ee
where for each $i$, $\hat{\imath}\in\{5,6,7,8\}$ and
$\hat{\imath}\ne \hat{\jmath}$ if $i\ne j$, one finds
\be
\{q_{\dot{a}},q_{\dot{b}}\}=2(N-\tilde{N}) \delta_{\dot{a}\dot{b}}
+{m \over 2}
\left( \gamma^{ji} \Pi M + M \Pi \gamma^{ij} \right)_{\dot{a}\dot{b}}
\left( J^{ij} \pm J^{\hat{\imath} \hat{\jmath}} \right) \,.
\ee
It is again clear that the rotations that appear on the right hand
side leave the world-volume of the $OD5$-brane invariant.

Finally, for the case of the $OD3$-brane, the relevant
generator of the supersymmetry algebra is not just $q$ (with $M=M_3$),
but rather $P_3^+ q$. The relevant subalgebra of the supersymmetry
algebra is then
\be\label{susyod3}
\{P_3^+ q_{\dot{a}}, P_3^+ q_{\dot{b}}\}
=2(N-\tilde{N})(P_3^+)_{\dot{a}\dot{b}}
+m (P_3^+ \g^{12}\Pi M_3)_{\dot{a}\dot{b}}(J^{12}-J^{56})\,.
\ee
The rotations that appear on the right hand side of (\ref{susyod3})
leave the world-volume of the $OD3$-brane invariant and act
trivially on the transverse space. This is again as expected since the
$OD3$-brane can be moved along the transverse directions $x^I$
with $I=3,4,7,8$, and thus is in general not invariant under any
rotation of these directions.
%%%%

\subsection{Comparison with results based on $(2,2)$
supersymmetry}

Finally, we explain how our results tie in with the
results of \cite{hk}. In \cite{hk} only the $(2,2)$ world-sheet
supersymmetry of the background was considered, and thus it was only
assumed that there are two complex Killing spinors (that were denoted
by $|0\rangle$ and $|\tilde{0}\rangle$ in \cite{hk}). In terms of our
description, these two states are the two spinor states that transform
in an irreducible   2-dimensional representation of the diagonal
$SO(4)$ subgroup of the background symmetry (see \cite{gg} for more
details). Let us denote by $\psi$ and $\chi$ the two left-moving
components, and by $\tilde\psi$ and $\tilde\chi$ their right-moving
counterparts. (In terms of the notation of \cite{gg}, $\chi$ and
$\tilde\chi$ are the `top', and  $\psi$ and $\tilde\psi$ the `bottom'
states.)
We now propose that the relation between the two sets of states is
given by
\ba
|0\rangle & = &
\psi - \tilde{\psi} + i (\chi + \tilde{\chi}) \,, \\
|\tilde{0}\rangle & = &
\chi - \tilde{\chi} + i (\psi + \tilde{\psi}) \,.
\ea
The main evidence for this proposal is that it correctly reproduces
the unbroken supersymmetries for the different branes that were
considered by \cite{hk}. For example, it was shown in \cite{hk} that
the supersymmetries that are preserved by the $OD3$-brane are
proportional to $|\tilde{0}\rangle$. Given the above identification,
this means that the preserved supersymmetries should be proportional
to $\chi-\tilde{\chi}$ and $\psi + \tilde{\psi}$. This agrees with
our analysis above since the preserved dynamical supersymmetry
generators include the `top' and `bottom' component of
$Q+iM_3\tilde{Q}$. In fact, on the space spanned by the top and bottom
component $M_3$ takes the form
\be\label{M3thematrix}
M_3 =\pmatrix{i& 0 \cr 0&-i}\,,
\ee
and thus we have $\psi=\tilde{\psi}$ and $\chi=-\tilde{\chi}$ at the
boundary.

Similarly, for the case of the $OD5$-brane, the unbroken
supersymmetries of \cite{hk} are
$\alpha |0\rangle + \zeta |\tilde{0}\rangle$, where
$\alpha=-i\zeta^*$. The surviving supersymmetries are thus
\be\label{survive}
2 \zeta_1 ( \chi+i\tilde{\psi}) - 2 \zeta_2 (\psi + i \tilde{\chi})
\ee
On the space spanned by the top and bottom component $M_5$ takes the
form
\be
M_5 = \pmatrix{-1 & 0 \cr 0 & -1} \,,
\ee
and thus we have $\psi=i\tilde{\psi}$ and $\chi =i\tilde{\chi}$ at
the boundary, in agreement with (\ref{survive}) above.

%%%%%%%%%%%%%%%%%%%%%%%

\section{The curved $D7$-brane}

The branes we have discussed so far all have world-volumes that are
flat hyperplanes in the transverse space. As was explained in
the introduction, one expects on general grounds that any world-volume
along which the superpotential (\ref{supco}) is constant should define a
supersymmetric brane. The simplest example is obviously the
brane whose world-volume is {\it defined} by (\ref{conscon}) with $W$
given by (\ref{supco}). In terms of real coordinates, this
world-volume is characterised by the two equations\footnote{In the
following we shall always choose the convention that $i,j$ run from
$1,\ldots 4$, while $I,J$ run from $1,\ldots, 8$.}
\be\label{const1}
\sum_{i=1}^{4} ( x^i x^i - x^{i+4} x^{i+4} ) = c_1 \,, \qquad
\hbox{and}  \qquad
\sum_{i=1}^4 x^i x^{i+4} = c_2 \,.
\ee
Since there are two `Dirichlet' directions (whose orientation depends
on the position on the world-volume of the brane), the corresponding
brane is a $D7$-brane \cite{hk}. In this section we want to analyse
the open string whose endpoint lies on this hypersurface. In
particular, we shall show that, together with suitable boundary
conditions for the fermions, this boundary condition is invariant
under two (dynamical) supersymmetry transformations.

The first step of the construction is to identify the matrix
$M$ that describes the combination of supersymmetries that should be
preserved by the brane. So far, we have always taken $M$ to be the
product of the $\gamma$-matrices associated to the Neumann directions,
but we could have equally taken $M$ to be the product of the
$\gamma$-matrices associated to the Dirichlet directions. For the
problem at hand this is more convenient, and we therefore define
\be\label{M7ansatz}
M_7 = {1\over x^I x^I}
         \left( x^i \gamma^i - x^{i+4} \gamma^{i+4} \right)
         \left( x^i \gamma^{i+4} + x^{i+4} \gamma^i \right) \,.
\ee
It is not difficult to check that $M$ is orthogonal, and that
\be\label{prop1}
M^t = - M \,.
\ee

For the following analysis it is convenient to work with a more
explicit description of the spinor representations. Let us write the
$\gamma$-matrices in terms of fermionic creation and annihilation
operators,
\ba\label{gammamatrices}
\gamma^i & = & b^i + (b^i)^\dagger \,, \\
\gamma^{i+4} & = & i \left(b^i - (b^i)^\dagger\right) \,,
\ea
where $\{b^{\dagger\, i}, b^j\} = \delta^{ij}$.
The top and bottom components of a spinor (that will be denoted by
$\uparrow$ and $\downarrow$ in the following) are the states that are
annihilated by all $(b^i)^\dagger$ or all $b^i$, respectively. These
are the states with weights $(+,+,+,+)$ and $(-,-,-,-)$.  The dotted
spinor states have an even number of $b$ excitations and thus
contain the top and bottom components as well as six other components
of the form $(+,+,-,-)$, together with permutations of $+$ and $-$.
Undotted spinor states are created from these states by the action of
an odd number of creation or annihilation operators.  An undotted
spinor therefore consists of two types of states: one of these has
weights $(+,+,+,-)$ together with three other permutations,  and can
be written as $|i \uparrow\rangle$ where $i$ labels the four distinct
states. The other type of state has $(-,-,-,+)$ and permutations with
one $+$, and can be written as $|i\downarrow\rangle$.  It will prove
convenient to decompose the $SO(8)$ spinors into these two types of
components.  For example, we will write the eight components of $S$ as
$S = (S^{i\, \uparrow},S^{i\,\downarrow})$, and likewise for
$\tilde{S}$. Note that
$\{S^{i\, \uparrow}, S^{j\, \downarrow}\} =\delta^{ij}$ and
$\{S^{i\, \uparrow}, S^{j\,\uparrow}\} =0$.
It is also convenient to introduce the following complex combinations
of left- and right-moving fermionic fields
\ba
\calT^a & = & \tilde{S}^a + i S^a \,, \label{Tdef}\\
\bar \calT^{a} & = & \tilde{S}^{a}  - i S^{a} \,, \label{Tbdef}
\ea
where $a=i\uparrow$ or $a=i\downarrow$. Finally, for the bosonic
coordinates we introduce the complex coordinates\footnote{We reserve
upper-case letters for fields, and use lower-case letters to describe
the geometry.}
\be
Z^i = X^i + i X^{i+4} \,, \qquad \bar Z^i = X^i - i X^{i+4} \,.
\ee
In these coordinates, $M_7$ then takes the form
\be
M_7=-i \left[ b^i,(b^\dagger)^j \right]
\frac{Z^j\bar{Z}^i}{|Z|^2}\,. \label{M7complex}
\ee
It is easy to see that $M_7$ acts on the top- and bottom components of
the (dotted) spinors precisely as $M_3$ in (\ref{M3thematrix}); this
is in agreement with the result of \cite{hk}. Under the dynamical
supersymmetry transformations associated to
$\epsilon_2=\uparrow$ or $\epsilon_2=\downarrow$ with
$\epsilon_1=-M_7\epsilon_2$ the fields then transform as
\be\label{bosonvar}
\begin{array}{lcllcl}
\delta_\uparrow Z^i & = & 0 &
\delta_\downarrow Z^i & = & 2\, \calT^{i\,\uparrow}  \\[4pt]
\delta_\uparrow \bar Z^i & = & 2\, \bar\calT^{i\,\downarrow}
\qquad &
\delta_\downarrow \bar Z^i & = & 0 \,,
\end{array}
\ee
and
\be\label{fermionvar}
\begin{array}{lcllcl}
\delta_\uparrow \calT^{i\,\uparrow} & = & (\partial_+ +\partial_-) Z^i
\qquad &
\delta_\downarrow  \calT^{i\,\uparrow} & = & 0  \\[4pt]
\delta_\uparrow \bar\calT^{i\,\uparrow} & = & (\partial_+ -\partial_-)
Z^i &
\delta_\downarrow  \bar\calT^{i\,\uparrow} & = & -2i\hm\bar Z^i  \\[4pt]
\delta_\uparrow \calT^{i\,\downarrow} & = &  2 i \hm Z^i &
\delta_\downarrow  \calT^{i\,\downarrow} & = &
(\partial_+ -\partial_-) \bar Z^i \\[4pt]
\delta_\uparrow \bar\calT^{i\,\downarrow} & = & 0 &
\delta_\downarrow  \bar\calT^{i\,\downarrow} & = &
(\partial_+ +\partial_-) \bar Z^i  \,,
\end{array}
\ee
as follows directly from (\ref{dyn}).

\subsection{The supersymmetry algebra}

The above transformation formulae must obviously respect the relations
that arise from the supersymmetry algebra. In particular, one easily
sees that
\be\label{crucial1}
\delta_\uparrow \delta_\uparrow \equiv 0 \equiv
\delta_\downarrow \delta_\downarrow  \,.
\ee
Furthermore, the anti-commutator of $\delta_\uparrow$ and
$\delta_\downarrow$ must satisfy
\be\label{crucial2}
\{\delta_\uparrow,\delta_\downarrow\} \equiv 2
(\partial_+ + \partial_-) = 2 \partial_\tau \,.
\ee
It is easy to see that (\ref{crucial2}) holds when acting on the
bosonic fields, or $\calT^{i\uparrow}$ and $\bar\calT^{i\downarrow}$,
but the analysis for the other two fermionic components requires more
care:
\ba
\{\delta_\uparrow,\delta_\downarrow \} \bar\calT^{i\uparrow} & = &
\delta_\downarrow (\partial_+ - \partial_-) Z^i
- 2 i \hm  \delta_\uparrow \bar Z^i \\
& = & 2 \left[ (\partial_+ - \partial_-) \calT^{i\uparrow}
 - 2 i \hm \bar\calT^{i\downarrow} \right]  \\
& = & 2 \left[ (\partial_+ + \partial_-) \bar\calT^{i\uparrow}
+ 2 i \partial_+ S^{i\uparrow} - 2 \partial_- \tilde{S}^{i\uparrow}
- 2 i \hm \bar\calT^{i\downarrow} \right] \,.
\ea
The last three terms in the bracket vanish by virtue of the equations
of motion
\ba
\partial_+ S^{i\uparrow} & = & \hm \Pi \tilde{S}^{i\uparrow} =
\hm \tilde{S}^{i\downarrow} \\
\partial_- \tilde{S}^{i\uparrow} & = &
- \hm \Pi S^{i\uparrow} = - \hm S^{i\downarrow} \,,
\ea
which imply, in particular, that
\be
2 i \partial_+ S^{i\uparrow} - 2 \partial_- \tilde{S}^{i\uparrow} =
2 i \hm \left( \tilde{S}^{i\downarrow} - i S^{i\downarrow} \right)
= 2 i \hm \bar\calT^{i\downarrow} \,.
\ee
The analysis for the $\calT^{i\downarrow}$ components is similar.

\subsection{The supersymmetry analysis}

We are now ready to analyse the supersymmetry of the boundary
conditions that characterise the curved $D7$-brane. For the bosonic
fields the `Dirichlet' boundary conditions are
\be \label{bosonD71}
Z^i Z^i = c \,,\qquad \hbox{and} \qquad
\bar Z^i \bar Z^i = \bar c \,,
\ee
where $c$ is a complex constant (and $\bar c$ its complex conjugate).
Furthermore, we postulate that the fermionic boundary conditions are
given by
\be
S =-M_7 \tilde{S} \,,
\ee
where $M_7$ is defined by either (\ref{M7ansatz}) or equivalently by
(\ref{M7complex}). In terms of the fermion components we have
introduced above, this boundary condition can be rewritten as
\be
Z^i \calT^{i\uparrow} = 0 \,, \qquad
\bar Z^i \bar\calT^{i\downarrow} = 0 \label{Fermion12} \,,
\ee
as well as
\be
P^{\bar\jmath l} \bar\calT^{l\uparrow} = 0 \,, \qquad
P^{\bar\jmath l} \calT^{j\downarrow} = 0 \,. \label{Fermion34}
\ee
Here $P^{\bar\jmath l}$ is the projection matrix defined by
\be
P^{\bar\jmath l}= \left( \delta^{jl} - {\bar Z^j Z^l \over |Z|^2}
\right) \,.
\ee
For future reference we note that
\be
\delta_\uparrow P^{\bar\jmath l} =
- 2{Z^l \over |Z|^2} P^{\bar\jmath r} \bar\calT^{r\downarrow}  \,,
\qquad
\delta_\downarrow P^{\bar\jmath l} =
- 2{\bar Z^j \over |Z|^2} P^{\bar r l} \calT^{r\uparrow} \,.
\label{Pplusminus}
\ee

It is easy to see that the variations of (\ref{bosonD71}) with respect
to $\uparrow$ and $\downarrow$ lead to the fermionic boundary
conditions (\ref{Fermion12}). The variation of these fermionic
boundary conditions then either vanish or give
\be
Z^i \partial_\tau Z^i = {1\over 2}\,
\partial_\tau  \left( Z^i Z^i \right) = 0 \,,
\ee
as follows from (\ref{crucial1}) and (\ref{crucial2}), or by direct
calculation. These two sets of boundary conditions are therefore
invariant under the two supersymmetry transformations. This leaves us
with analysing the supersymmetry transformations of
(\ref{Fermion34}). Using (\ref{Pplusminus}) one calculates
\be
\delta_\downarrow \left(P^{\bar\jmath l} \bar\calT^{l\uparrow} \right)
= -2 {\bar Z^j \over |Z|^2} P^{\bar r l} \calT^{r\uparrow}
\bar\calT^{l\uparrow} - 2 i \hm P^{\bar\jmath l} \bar Z^l \nonumber
= 0 \label{important1}
\ee
since
\be
 - {\bar Z^j \over |Z|^2} P^{\bar r l} \calT^{r\uparrow}
\bar\calT^{l\uparrow} = - {\bar Z^j \over |Z|^2} \calT^{r\uparrow}
P^{\bar r l}\bar\calT^{l\uparrow} = 0
\ee
by (\ref{Fermion34}), and
\be
P^{\bar\jmath l} \bar Z^l = \left( \delta^{jl} - {\bar Z^j Z^l \over |Z|^2}
\right) \bar Z^l = \bar Z^j - \bar Z^j {|Z|^2 \over |Z|^2} = 0 \,.
\ee
Likewise, one shows that
\be\label{important2}
\delta_\uparrow \left(P^{\bar\jmath l} \calT^{j\downarrow} \right) =
0 \,.
\ee
On the other hand, one finds
\be \label{Neumann1}
\delta_\uparrow \left(P^{\bar\jmath l} \bar\calT^{l\uparrow} \right)
= P^{\bar\jmath l} \partial_\sigma Z^l
+ 2{Z^l \bar\calT^{l\uparrow} \over |Z|^2}
P^{\bar\jmath r} \bar\calT^{r\downarrow}
\ee
and
\be \label{Neumann2}
\delta_\downarrow \left(P^{\bar\jmath l} \calT^{j\downarrow} \right)
= P^{\bar\jmath l} \partial_\sigma \bar Z^j
+2 {\bar Z^j \calT^{j\downarrow} \over |Z|^2}
P^{\bar r l} \calT^{r\uparrow} \,.
\ee
Apart from the bilinear fermion term, the right hand sides of
(\ref{Neumann1}) and (\ref{Neumann2}) are the Neumann boundary
conditions one would have expected. Since the bilinear fermion term
does not vanish automatically, we are therefore led to postulate
that the {\it actual (Neumann) boundary conditions} for the bosons are
given by
\ba
P^{\bar\jmath l} \partial_\sigma Z^l
+ 2{Z^l \bar\calT^{l\uparrow} \over |Z|^2}
P^{\bar\jmath r} \bar\calT^{r\downarrow}  & = & 0 \nonumber\\
P^{\bar\jmath l} \partial_\sigma \bar Z^j
+2 {\bar Z^j \calT^{j\downarrow} \over |Z|^2}
P^{\bar r l} \calT^{r\uparrow} & = & 0 \,. \label{Neumann12}
\ea
These two equations are complex conjugates of one another. As we shall
see below, they can also be obtained by analysing the variation of the
action directly.

Given the supersymmetry relations (\ref{crucial1}) and
(\ref{crucial2}), as well as (\ref{important1}) and (\ref{important2})
it is now easy to see that (\ref{Fermion34}) and (\ref{Neumann12}) are
closed under the different supersymmetry transformations. For example,
(\ref{crucial1}) and (\ref{important1}) imply that
\be
\delta_\uparrow \left(
P^{\bar\jmath l} \partial_\sigma Z^l
+ 2{Z^l \bar\calT^{l\uparrow} \over |Z|^2}
P^{\bar\jmath r} \bar\calT^{r\downarrow}\right) = 0 \,,
\ee
and in fact this is not difficult to check explicitly.
Furthermore, (\ref{crucial2}) and (\ref{important1}) imply that
\be
\delta_\downarrow \left(
P^{\bar\jmath l} \partial_\sigma Z^l
+ 2{Z^l \bar\calT^{l\uparrow} \over |Z|^2}
P^{\bar\jmath r} \bar\calT^{r\downarrow}\right) =
2 \partial_\tau \left( P^{\bar\jmath l} \bar\calT^{l\uparrow} \right) \,,
\ee
and thus the right hand side vanishes because of
(\ref{Fermion34}). The other variations are similar.

\subsection{Varying the action}

As mentioned above, the modified bosonic boundary conditions
(\ref{Neumann12}) can also be obtained directly by varying the
action.\footnote{Related modifications of the Neumann conditions by
fermion-bilinears have recently also been obtained in \cite{Lindstrom}.}
The boundary terms in the variation of the action $I$ are of the form
\be\label{variation}
\delta I_{bound} = \int\,d\tau\left(\delta X^{\mu}
\frac{\delta I}{\delta(\partial_\sigma X^\mu)}
+\delta S\frac{\delta I}{\delta(\partial_\sigma S)}
+\delta \tS\frac{\delta I}{\delta(\partial_\sigma \tS)}\right)\,,
\ee
where $\partial_\tau=\partial_+ + \partial_-$,
$\partial_\sigma =\partial_+ - \partial_-$. Using the explicit form of
the action (\ref{action}), the last two terms in the integrand are
\be
-\frac{1}{2}\left(S\delta S-\tS \delta \tS\right)\,.
\ee
This usually vanishes by imposing the boundary condition $S=-M\tS$,
provided that $M$ is an orthogonal constant matrix. In the present
context, however, $M$ depends on $Z^i$ and $\bar Z^i$, and $S=-M\tS$
therefore leads to
\be
\frac{1}{2}\left[
\delta Z^i \left(S \frac{\partial M}{\partial Z^i} \tS\right)
+\delta \bar{Z}^i \left(S \frac{\partial M}{\partial \bar{Z}^i}
\tS\right)\right]\,. \label{addterm}
\ee
On the other hand, the first term in (\ref{variation}) leads to
\be
-\frac{1}{4}\left(\delta Z^i\, \partial_\sigma \bar{Z}^i
+\delta \bar{Z}^i\, \partial_\sigma Z^i\right) \,,
\ee
where
$2\delta X^I\partial_\sigma X^I=(\delta Z^i \partial_\sigma \bar{Z}^i
+\delta \bar{Z}^i\partial_\sigma Z^i)$ has been used. Combining with
(\ref{addterm}) the variation of $Z^i$ and $\bar{Z}^i$ thus
has to satisfy
\be
\delta Z^i
\left(\frac{1}{2}\partial_\sigma\bar{Z}^i
- S \frac{\partial M}{\partial Z^i} \tS \right)
+\delta \bar{Z}^i \left(\frac{1}{2}\partial_\sigma Z^i
-S \frac{\partial M}{\partial \bar{Z}^i} \tS\right) = 0 \,.
\ee
The term proportional to $\delta Z^i$ can be written as
\be
\delta Z^i \left(\frac{1}{2} \left[ \delta^{ij}-\frac{\bar{Z}^j
Z^i}{|Z|^2} \right] \partial_\sigma \bar{Z}^j+\frac{\bar{Z}^j
Z^i}{2 |Z|^2}\partial_\sigma \bar{Z}^j-S \frac{\partial M}{\partial
Z^i} \tS\right)=0\,. \label{bcond}
\ee
The second term vanishes since $Z^i Z^i=c$ and the remaining terms
combine to give the Neumann conditions
\be
\delta Z^i \left(
P^{\bar{j}i}\partial_\sigma\bar{Z}^j-2 S \frac{\partial M}{\partial
Z^i} \tS \right) = 0 \,. \label{bilin}
\ee
Thus the Neumann conditions get indeed modified by a fermion bilinear
term.

One can readily verify that (\ref{Neumann12})
satisfies (\ref{bilin}). In order to see this, one observes that
(\ref{Fermion12}) implies
\be
\frac{\bar{Z}^l \calT^{l\downarrow}}{|Z|^2}
=2i\frac{\bar{Z}^k}{|Z|^2} S^{k\downarrow} \,,
\ee
while (\ref{Fermion34}) gives
\be
P^{j\bar r}\calT^{r\uparrow}=2i P^{j\bar r}P^{k \bar r}S^{k\uparrow} \,.
\ee
Putting these together one finds
\be
\frac{\bar Z^k \calT^{k\downarrow}}{|Z|^2}
P^{\bar r i} \calT^{r\uparrow}
=-4\frac{{\bar Z}^k}{|Z|^2}S^{k\downarrow}
P^{i\bar r}P^{l \bar r}S^{l\uparrow}\,.
\ee
Furthermore $P^{i\bar r}P^{l\bar r}= P^{l\bar i} + Z^i V^l$, and since
$Z^i\delta Z^i=0$, (\ref{bilin}) equals
\be
\delta Z^i \left(
8\frac{\bar{Z}^k}{|Z|^2}S^{k\downarrow}
P^{\bar i l}S^{l\uparrow}
- 2 S \frac{\partial M}{\partial Z^i} \tS \right) = 0 \,.
\ee
This then vanishes manifestly since it follows from
(\ref{M7complex}) that
\be
\frac{\partial M}{\partial
Z^i}=2i\frac{\bar{Z}^j}{|Z|^2}P^{\bar{i}k}(b^\dagger)^k b^j \,.
\ee

%%%%

\subsection{The curved $D5$-brane}

The analysis for the curved $D7$-brane can
be generalised to other curved branes. For example, the superpotential
is constant on the world-volume of a curved $D5$-brane that is
characterised by the equations
\ba\label{D5}
Z_1^2+Z_2^2+Z_3^2 &=& a \label{c5Dirichlet1}\\
\qquad Z_4 &=& b \label{c5Dirichlet2} \,.
\ea
The fermionic gluing matrix, $M_{c5}$, can be taken to be the product
of the gamma-matrices in the Dirichlet directions, and is explicitly
given by
\be\label{M5}
M_{c5} = -i \frac{Z^a\bar{Z}^c}{Z^d \bar{Z^d}} \left[ b^c,
\left(b^a\right)^\dagger \right] \gamma^4 \gamma^8 \,.
\ee
Here the indices $a,b,\ldots$ run over the reduced range $1,2,3$.
On the top and bottom states, $M_{c5}=-\bbbone$, and thus the
corresponding dynamical supersymmetries are related as
$\epsilon_1 = \epsilon_2$. The analogues of (\ref{bosonvar}) is then
\ba
&\delta_\uparrow Z^i =0 \,,\qquad &  \delta_\downarrow \bar{Z}^i=0 \\
&\delta_\uparrow \bar{Z}^i = {\cal J}^{i\downarrow} \,,\qquad &
\delta_\downarrow Z^i = {\cal J}^{i\uparrow} \,,
\ea
where the relevant combinations of the fermions are now
\be
{\cal J}^a = \tilde{S}^a+S^a \,,\qquad \bar{\cal J}^a =
\tilde{S}^a-S^a \,.
\ee
Unlike the situation for the $D7$-brane the two expressions are are not
complex conjugates of each other. The variations of the fermions can
similarly be determined. The bosonic Dirichlet conditions follow
directly from (\ref{D5}), while  the fermionic conditions are again
given by $S=-M_{c5}\tS$. The modified Neumann boundary conditions are
in this case
\be\label{c5neumann}
\delta_{\uparrow} \left( \hat{P}^{a \bar{b}} \bar{\cal J}^{a\uparrow}
\right)
= \hat{P}^{c \bar b} \partial_\sigma Z^c
+ 2 \frac{Z^a\bar{\cal J}^{a\uparrow}}{Z^dZ^d}\,
\hat{P}^{e \bar b } \bar{\cal J}^{e\downarrow} = 0 \,,
\ee
as well as its complex conjugate, where the modified projector
$\hat{P}^{a\bar b}$ is defined by
\be\label{newP}
\hat{P}^{a\bar b} = \delta^{ab} - \frac{Z^a \bar{Z}^b}{Z^d\bar{Z}^d} \,.
\ee
As before, it can be shown that this set of boundary conditions is
invariant under the top- and bottom supersymmetry transformation. The
bilinear fermionic correction term to the bosonic Neumann condition
can also be derived from the action as before.

%%%%

\subsection{Curved branes and supersymmetry}

The above analysis implies that the curved $D7$-brane preserves two
dynamical supersymmetries (namely those associated to the top- and
bottom components), but one may wonder whether it preserves any more
supersymmetries. In order to analyse this question, it is convenient
to study first the corresponding question for the curved $D7$-brane
in flat space. The fact that this brane preserves at least two
supersymmetries follows directly from the above analysis (since the
whole argument goes through for $\hm=0$). In fact, the top- and bottom
components are the only constant eigenvectors of $M_7$, and one should
therefore expect that this brane does not preserve any additional
supersymmetries.

Curved supersymmetric branes of this type have been discussed before
from the point of view of calibrated geometry, \eg in
\cite{Gibbons}. In fact, it is known \cite{Harvey} that any complex
submanifold of $\Cop^n$ is calibrated, and thus allows for covariantly
constant spinors. If $F_i(z_1,\ldots,z_n)$, $i=1,\ldots, l$ is a
family of holomorphic functions, then the joint zero locus
\be
F_i(z_1, \cdots ,z_n) =0\,, \qquad i=1,\ldots, l
\ee
defines a complex submanifold. For the case of the curved
$D7$-brane, the world-volume is the deformed conifold (\ie, the
cotangent bundle of $S^3$, $T^* S^3$), which is a Calabi-Yau 3-fold
with holonomy group $SU(3)$. The theorem in \cite{Wang} then implies
that there are precisely $2$ covariantly constant spinors. Thus the
curved $D7$-brane in flat space preserves precisely the two
supersymmetries we have found above. In particular, this then implies
that the curved $D7$-brane in the pp-wave background also only
preserves these two supersymmetries (since the supersymmetry of a
brane cannot decrease by taking the limit $\hm\rightarrow 0$).

The analysis for the curved $D5$-brane is similar. By the same
arguments as above, the curved $D5$-brane also preserves two dynamical
supersymmetries in flat space. As regards the kinematical
supersymmetries, $M_{c5}$ acts as $+1$ on both $(+++-)$ and $(---+)$,
and therefore the curved $D5$-brane preserves also two kinematical
supersymmetries in flat space. (However, since
$\Pi M_{c5} \Pi M_{c5}=+1$ on both of these states, these do not give
rise to kinematical supersymmetries in the pp-wave background.) This
is consistent with the argument based on calibrated geometries. For
the curved $D5$-brane the world-volume is the cotangent bundle of
$S^2$, $T^*S^2$. This defines a Calabi-Yau 2-fold with holonomy group
$SU(2)$. The theorem in \cite{Wang} then implies again that there are
two dynamical supersymmetries. The additional kinematical
supersymmetries  arise from the additional two transverse flat
directions parametrised by $z_4=c$.
\medskip

The supersymmetry of the curved $D7$-brane is also consistent with the
following geometric picture of its world-volume: it can be thought of
as being a ruled surface, where the lines are given by the
world-volume of oblique $D3$-branes, which sweep out the $D7$-brane
under the action of the diagonal $SO(4)$ subgroup of the
$SO(4)\times SO(4)$ isometry of the background. More explicitly,
consider the oblique D3-brane whose world-volume is described by
(\ref{wv}) with $a=c^{1/2}$ and $b=0$. Then this D3-brane lies inside
the world-volume of the curved D7-brane that is described by the
equation
\be\label{D7world}
z_1^2 + z_2^2 + z_3^2 + z_4^2 = c \,.
\ee
Furthermore, any point on (\ref{D7world}) is related by an element in
the diagonal $SO(4)$ subgroup to a point on the above oblique
$D3$-brane. Thus the oblique $D3$-brane sweeps out the full
world-volume of the curved $D7$-brane under this $SO(4)$ action.
{}From the analysis in \cite{Douglasetal} it follows that the
different $D3$-branes, that are related by an $SO(4)$ rotation in
this manner, preserve two common supersymmetries, which are precisely
the top and bottom spinors in our context.

%%%%

\acknowledgments{
We are grateful to Costas Bachas, Gary Gibbons, Robert Helling and Nemani
Suryanarayana for useful conversations. MRG thanks the Royal Society
for a University Research Fellowship. SS-N is grateful to St John's
College, Cambridge, for a Jenkins Scholarship. AS is supported by the
Gates scholarship and the Perse scholarship of Gonville and Caius
College, Cambridge. }

%%%%


\begin{thebibliography}{99}

\bibitem{maldetal}{D.~Berenstein, J.~Maldacena, H.~Nastase,
{\it Strings in flat space and pp waves from ${\cal N}=4$ super Yang
Mills}, JHEP {\bf 0204}, 013 (2002); {\tt hep-th/0202021}.}

\bibitem{hulletal}{M.~Blau, J.~Figueroa-O'Farrill, C.~Hull,
G.~Papadopoulos, {\it Penrose limits and maximal supersymmetry},
Class.\ Quant.\ Grav.\ {\bf 19}, L87 (2002);
{\tt hep-th/0201081}.}

\bibitem{bp}{M.~Billo, I.~Pesando, {\it Boundary states for GS
superstrings in an $Hpp$ wave background}, Phys.\ Lett.\
{\bf B536}, 121 (2002); {\tt hep-th/0203028}.}

\bibitem{dp}{A.~Dabholkar, S.~Parvizi, {\it  Dp branes in pp-wave
background}, Nucl.\ Phys.\ {\bf B641}, 223 (2002);
{\tt hep-th/0203231}.}

\bibitem{st1}{K.~Skenderis, M.~Taylor, {\it Branes in AdS and pp-wave
spacetimes}, JHEP {\bf 0206}, 025 (2002); {\tt hep-th/0204054}.}

\bibitem{bgg}{O.~Bergman, M.R.~Gaberdiel, M.B.~Green, {\it $D$-brane
interactions in type IIB plane-wave background},
JHEP {\bf 0303}, 002 (2003); {\tt hep-th/0205183}.}

\bibitem{st2}{K.~Skenderis, M.~Taylor,
{\it Open strings in the plane wave background I: quantization and
symmetries}, {\tt hep-th/0211011}.}

\bibitem{gg}{M.R.~Gaberdiel, M.B.~Green, {\it The D-instanton and
other supersymmetric D-branes in IIB plane-wave string theory},
{\tt hep-th/0211122}, to appear in Ann. Phys.}

\bibitem{st3}{K.~Skenderis, M.~Taylor,
{\it Open strings in the plane wave background II: superalgebras and
spectra}, {\tt hep-th/0212184}.}

\bibitem{mm}{J.~Maldacena, L.~Maoz,
{\it Strings on pp-waves and massive two dimensional field theories},
JHEP {\bf 0212}, 046 (2002); {\tt hep-th/0207284}.}

\bibitem{hk}{Y.~Hikida, S.~Yamaguchi,
{\it D-branes in pp-waves and massive theories on worldsheet with
boundary}, JHEP {\bf 0301}, 072 (2003); {\tt hep-th/0210262}.}

\bibitem{hiv}{K.~Hori, A.~Iqbal, C.~Vafa, {\it D-Branes and mirror
symmetry}, {\tt hep-th/0005247}.}

\bibitem{met}{R.R.~Metsaev, {\it Type IIB Green-Schwarz superstring in
plane wave Ramond-Ramond background}, Nucl.\ Phys.\ {\bf B625}, 70
(2002); {\tt hep-th/0112044}.}

\bibitem{mt}{R.R.~Metsaev, A.A.~Tseytlin, {\it Exactly solvable model
of superstring in plane wave Ramond-Ramond background}, Phys.\ Rev.\
{\bf D65}, 126004 (2002); {\tt hep-th/0202109}.}

\bibitem{ggut}{M.B.~Green, M.~Gutperle, {\it Light-cone supersymmetry
and $D$-branes}, Nucl.\ Phys.\ {\bf B476}, 484 (1996);
{\tt hep-th/9604091}.}

\bibitem{Lindstrom}
{C.~Albertsson, U.~Lindstrom, M.~Zabzine,
{\it Superconformal boundary conditions for the WZW model},
JHEP {\bf 0305}, 050 (2003); {\tt hep-th/0304013}.}

\bibitem{Gibbons}{G.W.~Gibbons, G.~Papadopoulos, {\it Calibrations
and intersecting branes}, Commun.\ Math.\ Phys.\  {\bf 202}, 593
(1999); {\tt hep-th/9803163}.}

\bibitem{Harvey}{
F.R.~Harvey, {\it Spinors and calibrations}, Academic Press (1990),
New York.}

\bibitem{Wang}{
McKenzie Y.~Wang, {\it Parallel spinors and parallel forms},
Ann.\ Global Anal.\ Geom.\ {\bf 7}, 59 (1989).}

\bibitem{Douglasetal}{M.~Berkooz, M.R.~Douglas, R.G.~Leigh,
{\it Branes intersecting at angles},
Nucl.\ Phys.\ B {\bf 480}, 265 (1996);
{\tt hep-th/9606139}.}



\end{thebibliography}
\end{document}